\pdfoutput=1
\RequirePackage{ifpdf}

\documentclass{JINST}
\usepackage{cite}

\title{A Measurement of the Absorption of Liquid Argon Scintillation Light
by Dissolved Nitrogen at the Part-Per-Million Level}

\author{B.J.P. Jones$^a$\thanks{Corresponding Author}, C.S. Chiu$^a$, J.M. Conrad$^a$, C. M. Ignarra$^a$, T. Katori$^a$, M. Toups$^a$.\\
\llap{$^a$}Massachusetts Institute of Technology,\\
  77 Massachusetts Avenue, Cambridge, MA 02139, United States of America\\
  E-mail: \email{bjpjones@mit.edu}}

\abstract{We report on a measurement of the absorption length of scintillation
light in liquid argon due to dissolved nitrogen at the part-per-million (ppm)
level. We inject controlled quantities of nitrogen into a high
purity volume of liquid argon and monitor the light yield from an alpha source.
The source is placed at different distances from a cryogenic 
photomultiplier tube assembly.  By comparing the light yield from each position we extract
the absorption cross section of nitrogen. We
find that nitrogen absorbs argon scintillation light with strength of 
$\left(1.51\pm 0.15\right)\times10^{-4} \;\mathrm{cm^{-1} ppm^{-1}}$, corresponding
to an absorption cross section of $\left(7.14\pm0.74 \right)\times10^{-21}\;\mathrm{cm^{2}~ molecule^{-1}}$.
We obtain the relationship between absorption length and nitrogen
concentration over the 0 to 50  ppm range and discuss the implications
for the design and data analysis of future large liquid argon time projection chamber (LArTPC)
detectors.  Our results indicate that for a current-generation LArTPC, where
a concentration of 2 parts per million of nitrogen is expected, the attenuation
length due to nitrogen will be $30 \pm 3$ meters.}

\keywords{Noble-liquid detectors; Photon detectors for UV, visible and IR photons; Scintillators, scintillation and light emission processes}

\begin{document}

\section{The Effects of Nitrogen upon Argon Scintillation Light}

Liquid argon scintillation light is used as a particle detection tool
in many current and future neutrino and dark matter experiments \cite{Akiri:2011dv, Rubbia:2012jr, Jones:2011ci, Menegolli:2012jq, Boulay:2012hq, Rielage:2012zz}. The features of liquid argon
which make it particularly apt as a detection medium in such experiments
include its relatively low cost, its high scintillation yield of tens
of thousands of photons per MeV, and the transparency of pure liquid
argon to its own scintillation light, which is emitted at a wavelength
of 128 nm \cite{Lippincott:2008ad}. Highly purified liquid argon can also achieve a very long
free-electron lifetime \cite{Baibussinov:2009gs}, making it an ideal active medium for liquid
argon time projection chamber (LArTPC) detectors. These detectors may utilize scintillation
light as a trigger, a cosmic-ray rejection tool, and a method to obtain
extra information for augmenting TPC-based reconstruction techniques.

It is widely known that whilst high-purity argon possesses
these useful qualities, tiny concentrations of impurities, such as
nitrogen at the part-per-million (ppm) level \cite{Acciarri:2008kx}  or oxygen and water at
the tens of parts-per-billion (ppb) level \cite{Acciarri:2008kv}, can have a very damaging
effect both upon the argon scintillation yield through the process
of scintillation quenching and upon the argon transparency at 128 nm
through ultraviolet absorption.  Argon scintillation proceeds
via the production of singlet and triplet eximer states, which subsequently decay to
 individual argon atoms and a 128 nm photon \cite{Suzuki:1979km}.  The singlet and triplet eximers
 have lifetimes of 6 ns and 1500 ns \cite{ICARUS_NIM} respectively,
which lead to argon scintillation light production with two characteristic time constants.
The quenching process involves an interaction of eximers with impurity
molecules, resulting in an excimer dissociation with no photon produced.  
Due to the longer lifetime, a triplet state is more likely to interact with an
impurity molecule before producing a scintillation photon than a singlet state. Hence
scintillation quenching typically affects the slow component of argon scintillation
light much more than the fast component, though both scintillation 
mechanisms can be significantly suppressed at large enough
concentrations.  Absorption, on the other hand, refers to the loss of emitted
128 nm photons by interaction with nitrogen molecules during propagation between production
and detection points, thus it affects the slow and fast components in the same way.  This is the quantity we aim to measure in this paper.

In LArTPC experiments, the concentrations of water and oxygen
must already be controlled at the tens to hundreds of parts-per-trillion (ppt) level
in order to maintain a long free-electron lifetime for charge drift \cite{Baibussinov:2009gs}. 
Nitrogen, however, does not damage the free-electron lifetime, and
is significantly more difficult to remove from the argon during the
purification process than oxygen or water.
Hence LArTPC experiments like MicroBooNE and LBNE expect to operate in
a regime with ppm levels of nitrogen dissolved in the active volume.  
Tight requirements upon the allowed nitrogen concentration to
prevent significant losses of scintillation light may be a cost
driver for the cryogenics system and active medium in such an experiment.

Previous studies of the quenching effects of nitrogen in small test
cells show that below around \textasciitilde{}2 ppm,
scintillation quenching from nitrogen is not likely to cause significant
problems for LArTPC experiments \cite{Acciarri:2008kv}. However, the effects 
of nitrogen absorption upon 128 nm argon scintillation light
travelling over long distances have not previously been measured. Knowledge of this vital
parameter is necessary to understand the argon purity requirements
for future large detectors.  Furthermore, it will be an important parameter for
detector simulations and data analysis of both current- and future-generation
LArTPC experiments with optical systems.

In this paper we describe measurements made using a high-purity liquid
argon test stand at the Fermi National Accelerator Laboratory (Fermilab).
 We inject controlled amounts
of nitrogen gas into high-purity argon and measure the light absorption
effect by comparing the light yield from a polonium-210 alpha source
held in two positions relative to a cryogenic photomultiplier
tube (PMT) assembly. From these measurements we are able to extract the
absorption cross section of dissolved nitrogen and explore its expected effects
in LArTPC detectors in the concentration range 0 - 50 ppm.

\section{Experimental Configuration}

The tests described in this paper were performed in the 220-liter
vacuum-insulated ``Bo'' cryostat at the Proton Assembly Building at Fermilab.
The cryostat is evacuated for several days to
remove contaminants before filling.  Then research-grade liquid argon is passed
through a system of regenerable filters and molecular sieves, described in 
\cite{Curioni:2009rt}, to fill the vessel with high-purity liquid argon
to a level between 76 and 89 cm. The remaining
few centimeters of the 55.9 cm diameter cylindrical vessel are occupied by argon vapor.

\begin{figure}[tb]
\begin{centering}
\includegraphics[width=1.0\columnwidth]{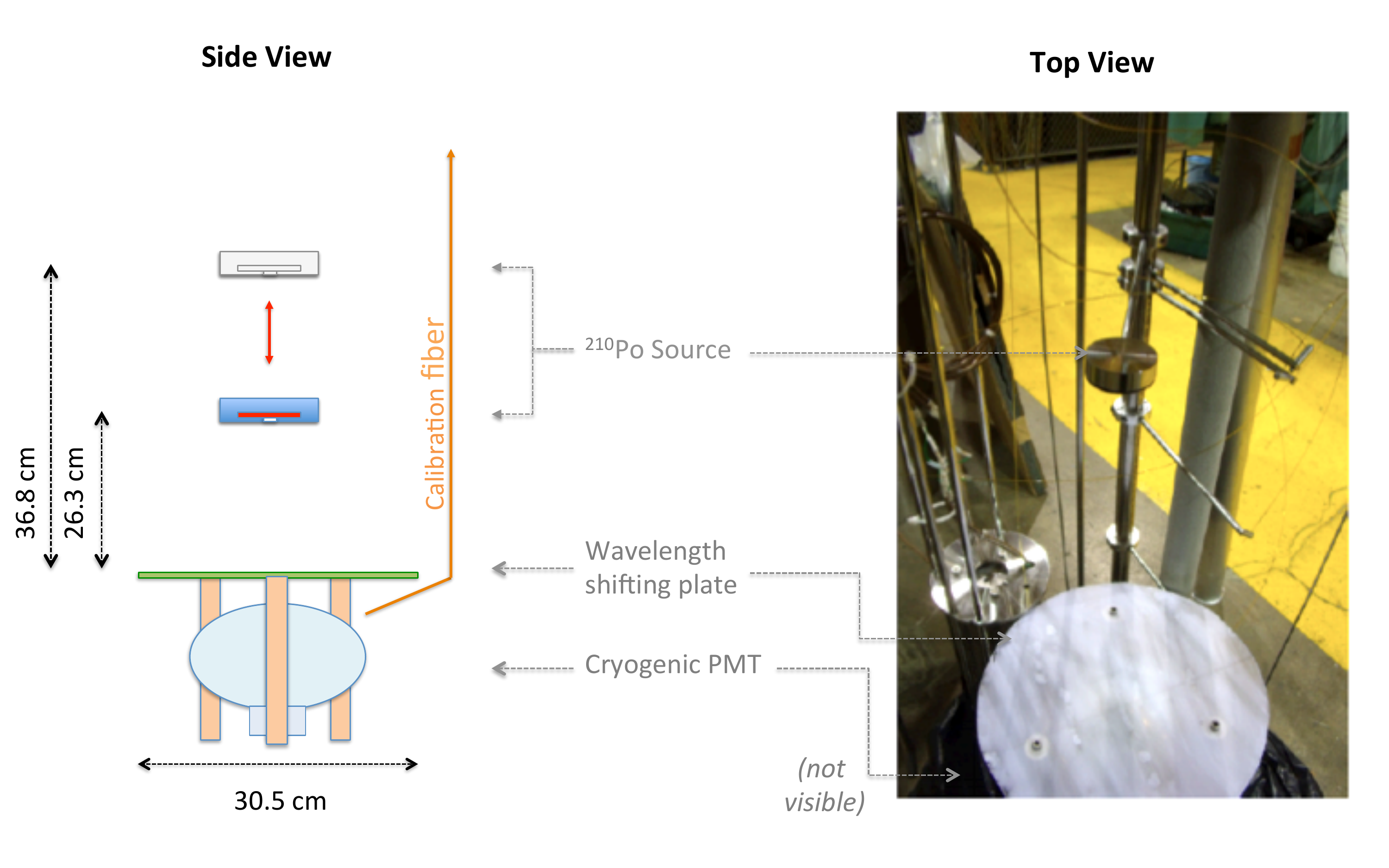}
\par\end{centering}

\caption{Diagram and photograph of the experimental configuration for this
study \label{fig:Diagram-and-photograph}}
\end{figure}

A diagram and photograph of the  apparatus inside the cryostat are shown 
in Figure \ref{fig:Diagram-and-photograph}.
 Attached to the cryostat lid is a support structure which holds a
single optical assembly consisting of a Hamamatsu R5912-02mod
cryogenic PMT, cryogenic base electronics, and a wavelength-shifting plate, more details about which can be found in \cite{Briese:2013wua, Chiu:2012ju, Jones:2012hm}. Argon scintillation light with a wavelength of
128 nm can be captured by tetraphenyl butadiene (TPB) in the plate
coating and re-emitted as visible light. The TPB emission spectrum
has a peak wavelength of around 450 nm \cite{Jones:2012hm}, which is well matched to the
peak quantum efficiency of the cryogenic photomultiplier tube \cite{HamamatsuPMT}. Some
of this light leads to the production of photoelectrons at the PMT
photocathode, which are amplified by the PMT dynode chain.
For this study the PMT high voltage is set at 1100 V, which provides
a stable gain of around $10^{7}$ \cite{Briese:2013wua}. Signal and high voltage are
both carried by one common 50$\Omega$ RG-58 cable which penetrates the cryostat through
a potted epoxy feedthrough \cite{EpoxyFT}. Outside the cryostat the AC signal is split
from the DC high voltage using a splitter unit, and the signal is
connected to a Tektronix DPO5000 oscilloscope terminated at 50 $\Omega$ and 
running at one gigasample per second. 

A polonium-210 disc source is held an adjustable distance above the
wavelength-shifting plate in a stainless steel source holder. This
source produces monoenergetic alpha particles of 5.3 MeV which scintillate
in the liquid argon and produce 128 nm scintillation light, which can
be detected by the PMT assembly. 

The number of photons detected for
each alpha particle is expected to be approximately Poisson distributed
with a mean determined by the initial photon yield, the geometrical
configuration of the experiment, and the assembly global collection
efficiency, which is defined as the number of amplified photoelectrons
produced for each incident 128nm photon on the wavelength-shifting plate. 
In this study we measure the light yield at different nitrogen concentrations
relative to a configuration with the same geometry and clean argon (defined below), thus making this 
analysis insensitive to the global collection efficiency of the assembly and the 
absolute scintillation yield, and only dependent upon the stability of the
photomultiplier gain and the stability of the scintillation yield.

The argon delivered to the Proton Assembly Building has a typical nitrogen concentration of $<$ 1 ppm.  It is then passed through molecular sieves
and filters to remove water and oxygen, a process which also removes some nitrogen.  We monitor nitrogen concentrations using an LD8000 Trace Nitrogen Analyzer \cite{LDetekAnalyzer}. The 
measured N$_2$ concentration in the argon delivered in a fresh fill of our  apparatus is 
37 ppb.  In this very low concentration range, our measurement accuracy suffers, since it is near the sensitivity limit
of our nitrogen monitor.  However, the nitrogen concentration is far below the 
range where the effects of quenching are significant \cite{Acciarri:2008kv}, and our results will show that it is also below the concentration range where we would expect
any significant absorption.  Hence this argon is "clean" for the purposes of our study.

The pressure inside the cryostat is maintained at 10 psi relative to atmospheric 
pressure with a liquid-nitrogen-cooled condenser tower, which re-condenses argon vapor into the liquid.
When the condenser is running, the cryostat is a closed system with
no vapor or liquid able to enter or leave the volume. The liquid level
is measured using a capacitive level monitor and for these studies
was stable at 79 cm, in contact with a 22 cm argon vapor region at
the top of the cryostat. 

We monitor trace impurity levels in both the liquid and gas phases
from capillary pipes which penetrate the cryostat lid and run
to oxygen, water, and nitrogen impurity monitors. Oxygen and water
 were both found to be present at the tens of ppb level.  
Since we perform measurements relative to an initial light yield, it is the 
stability of these impurity concentrations rather than their absolute
values which is important for our measurement.
The oxygen concentration was stable
to within 2 ppb, and the water concentration stable to within 4 ppb for both runs
documented in this paper.

\begin{figure}[tb]
\begin{centering}
\includegraphics[width=1\columnwidth]{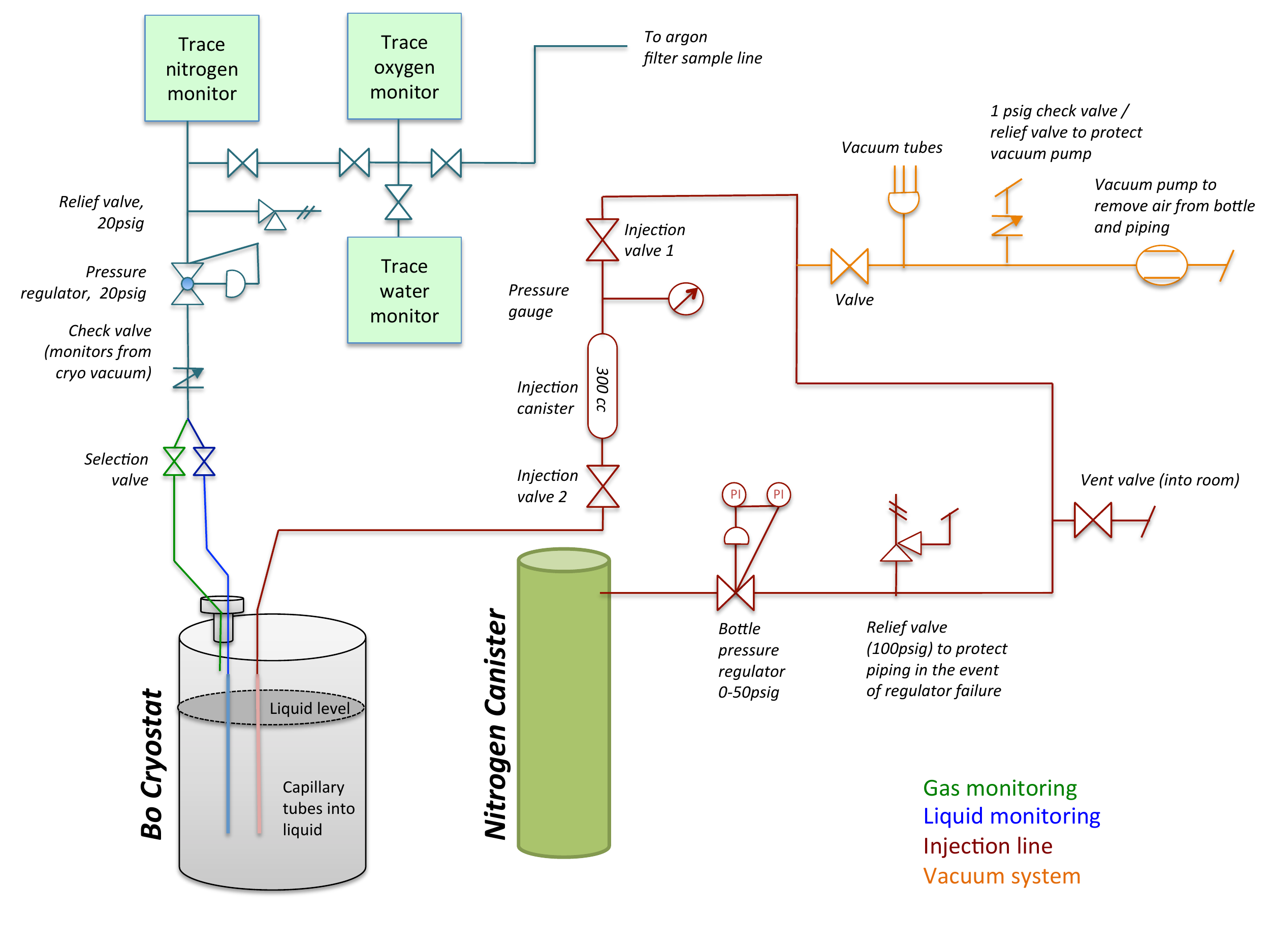}
\par\end{centering}

\caption{Schematic of the nitrogen injection and gas sampling apparatus used
in this test. \label{fig:Schematic-of-the}}
\end{figure}

We inject controlled amounts of gaseous nitrogen into
the liquid through an injection capillary. This capillary
is connected to a 300 $\mathrm{cm^3}$ canister which can be charged to a known
pressure with nitrogen from a gas cylinder which can then be released
into the liquid. After injection, the nitrogen inside Bo comes to a
vapor-liquid equilibrium state, with stable but different concentrations
in the liquid and gas phases. The equilibration process takes approximately
30 minutes. When not being charged for injection, the injection line
from the nitrogen cylinder up to and including the gas canister is
vented into the atmosphere and then evacuated 
to ensure no additional leakage of nitrogen into the volume occurs.  The complete nitrogen
injection and sampling apparatus is shown schematically in Figure \ref{fig:Schematic-of-the}.

This measurement uses the results of two experimental runs that were taken
approximately three weeks apart. This is the amount of time required
to disassemble, reconfigure, pump down and refill the apparatus. Each
run was started with 79 cm of high-purity argon, and nitrogen
was injected a few ppm at a time until we had reached
a concentration of approximately 50 ppm. After each set of injections
we measured the light yield visible from the alpha source. The two
runs had the source positioned at different distances from the PMT
assembly. We compare the detected light yield from a source placed
36.8 cm ("far configuration") from the assembly to the detected light yield from the same
source placed 20.3 cm ("near configuration") from the assembly as a function of nitrogen concentration.
Using this data we extract the absorption strength of nitrogen dissolved in argon.

In order to interpret the experimental data which will be presented,
first consider the effect of injecting a substance which leads to a particular
percentage of light absorption per centimeter and a simultaneous
scintillation quenching effect. If we were to inject a fixed quantity
of this substance and measure the same fractional light loss for both source configurations, 
we would conclude that the scintillation yield has been reduced and
 we have observed a pure quenching
process. If there is a greater fractional light loss for the
further source, we may be seeing either pure absorption, or a combined
quenching and absorption process. Taking the ratio of the remaining fractional light yield
from the near configuration to the remaining fractional light yield from the
far configuration, the quenching effect cancels, and we are left with a measurement only sensitive to the absorption strength. This is the quantity we aim to measure in this paper.  The effects of nitrogen upon scintillation
quenching of prompt and slow light have been studied elsewhere \cite{Acciarri:2008kv}, and
will be the subject of a future study with this apparatus \cite{FutureStudy}.

\begin{figure}[tb]
\begin{centering}
\includegraphics[width=1\columnwidth]{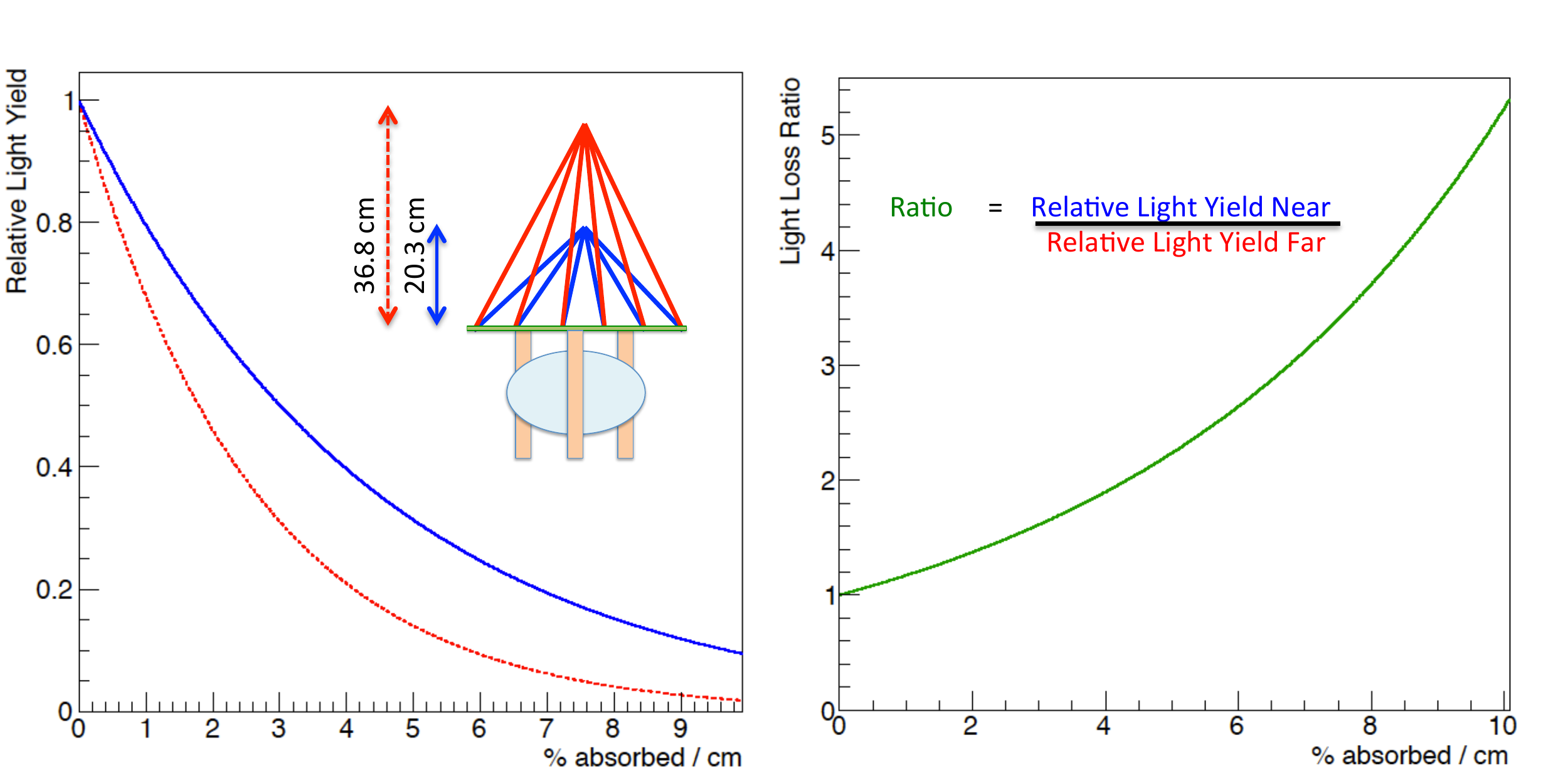}
\par\end{centering}

\begin{centering}
\caption{Left : Fractional light yield expected as a function of absorption strength
for each source position. Right: Ratio of the two fractional light yields,
in which any quenching effect cancels. \label{fig:RelLightYieldModel}}

\par\end{centering}

\end{figure}

The photons which propagate from source to plate have a distribution
of path lengths, as illustrated in the cartoon in Figure \ref{fig:RelLightYieldModel},
left. In order to understand the relative fractional light yield expected
between the two source positions for a given absorption strength,
we perform a ray tracing simulation and attenuate each ray according
to its distance of travel from source to plate.  The fractional light yield
at each source position due to absorption is shown in Figure \ref{fig:RelLightYieldModel}, left. We define the fractional light yield ratio as the fractional light yield from the near source position divided by the fractional light yield from the far source position.  In the presence of absorption this number is larger than one, as the light from the far source position is more strongly attenuated due to the longer path lengths of the light rays.  The fractional light yield ratio as a function of absorption strength extracted from the ray tracing model is shown in Figure \ref{fig:RelLightYieldModel}, right. 

Our goal is to find the conversion factor between
percentage absorption per centimeter and nitrogen concentration in argon by
comparing the measured fractional light yield ratio as a function of nitrogen concentration to the curve of Figure \ref{fig:RelLightYieldModel},
right.  We find that, in our region of interest of 0\% to 1\% per centimeter absorption,
corresponding to the concentration range 0 to 50 ppm,
the curve of Figure \ref{fig:RelLightYieldModel}, right can be well approximated by a 
straight line.  Thus we can extract the absorption strength of nitrogen by measuring the gradient of the linear relationship between fractional light yield ratio and nitrogen concentration, and from this also the molecular absorption cross section of dissolved nitrogen in argon to 128 nm light.

\section{Data Acquisition and Analysis}

\begin{figure}[tb]
\begin{centering}
\includegraphics[width=0.8\columnwidth]{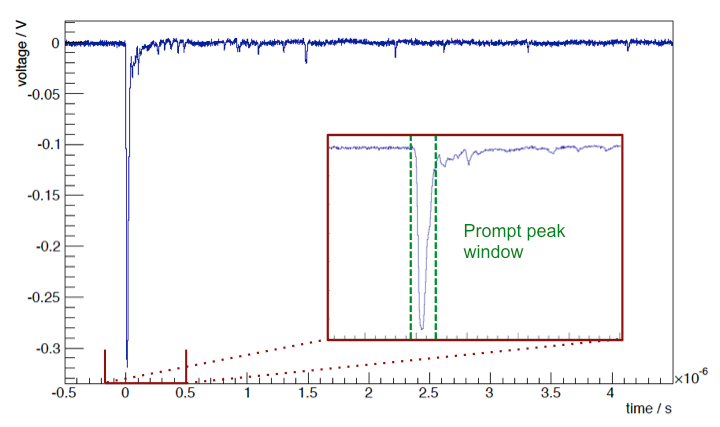}
\par\end{centering}

\caption{A typical PMT waveform produced by the light from a scintillating
alpha particle in the near source configuration. \label{fig:A-typical-PMT}}
\end{figure}

A typical PMT pulse initiated by alpha scintillation is shown in Figure \ref{fig:A-typical-PMT},
and the characteristic singlet and triplet scintillation components
of liquid argon with time constants 6 ns and 1500 $\mu$s respectively
are clearly visible as a large prompt peak followed by trailing
single photoelectron pulses, respectively. As described in the introduction, 
the slow light component from the triplet state is much more susceptible to scintillation quenching
than the prompt light component. Therefore, since
the focus of this study is nitrogen absorption, we consider only the
light in the prompt peak, measuring the total charge recorded between
-10 and +40 ns relative to the time of a $-30$ mV falling edge trigger.
We tested for the effects of pulse size bias due to this narrow prompt light window 
by comparing the pulse area spectrum recorded using this 50 ns window to the
pulse area distribution recorded using an extended 70 ns winow and observed no biasing effect.

\begin{figure}[tb]
\begin{centering}
\includegraphics[width=0.7\columnwidth]{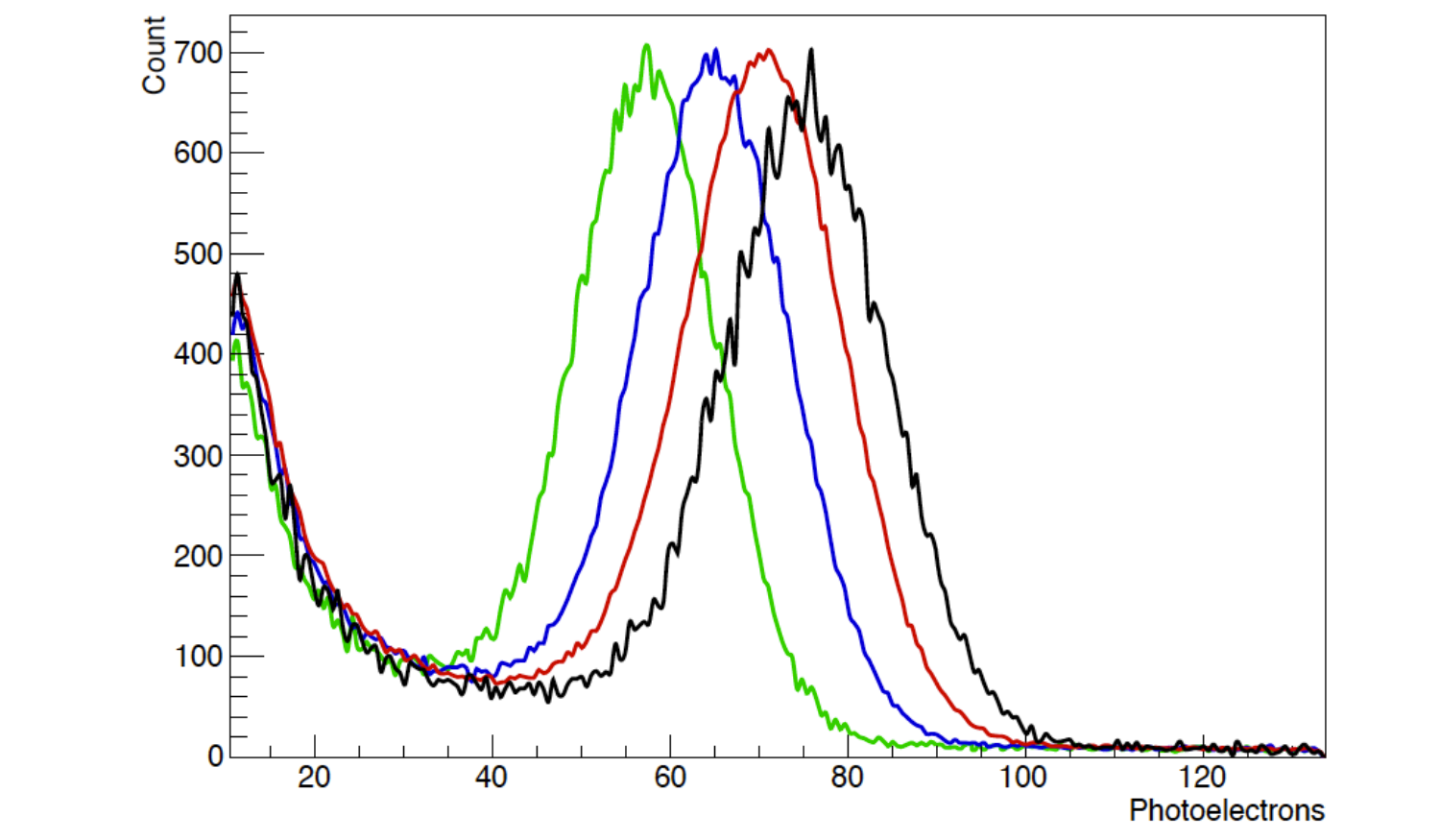}
\par\end{centering}

\caption{Pulse area distributions for the near configuration for, right to left (color online): 37 ppb
(black), 3.7  ppm (red), 7.4 ppm (blue), 15.5 ppm (green) of nitrogen
in argon.
\label{fig:SeveralDistributions}}
\end{figure}

To obtain the relative light yield detected at the PMT for each concentration point, PMT waveforms are recorded using a Tektronix DPO5000 oscilloscope, and the area within the prompt window is histogrammed in real time.  For illustration, the  distribution of pulse areas 
for a few concentration points
is shown in Figure \ref{fig:SeveralDistributions}. The falling cosmic-ray-induced
background remains approximately unchanged whilst the Poisson-like alpha peak is diminished in intensity by the injection of nitrogen.  

In order to quantify the charge recorded
in a light pulse from an alpha particle in terms of amplified photoelectrons, the
average single photoelectron charge was measured using a pulsed LED.  The measurement of the absolute single photoelectron scale and gain stability will be discussed in section \ref{sec:SPESection}.  

To extract the relative light yield from the source at each concentration point, we fit to a function which has a term describing the falling cosmic background and a term describing alpha-induced Poisson-like peak.  The construction of this function will be discussed in section \ref{sec:FitFuncSection}.

\subsection {Measurement of the Single Photoelectron Scale and Stability}
\label{sec:SPESection}

We normalize all measured light yields to the single photoelectron scale, which we refer to hereafter as <SPE>.  For this study, <SPE> is measured using a pulsed LED between every set of nitrogen injections.  It will be shown in section \ref{sec:FitFuncSection} that our attenuation length measurement is relatively insensitive to the absolute <SPE> scale so long as it is constant for the duration of each run.  However, the stability of the gain between different points within the same run is crucial.  The gain stability is obtained by calculating the spread of <SPE> values within the run.  Since the <SPE> spread gives the leading systematic error in this study, we make several independent cross-checks of gain stability, which all give similar results and will be described below.

\begin{figure}[tb]
\includegraphics[height=7.2cm]{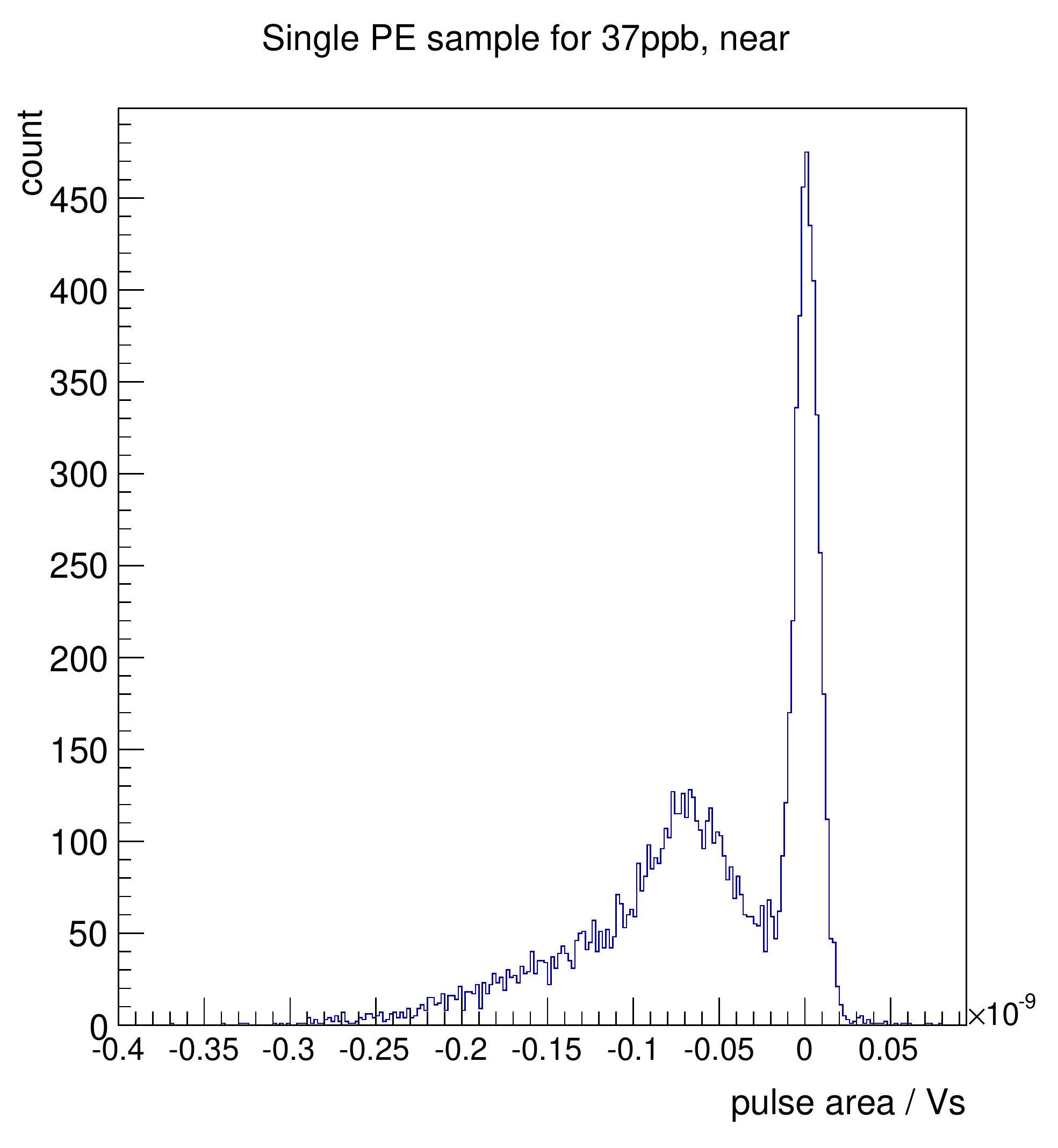}\includegraphics[height=7.0cm]{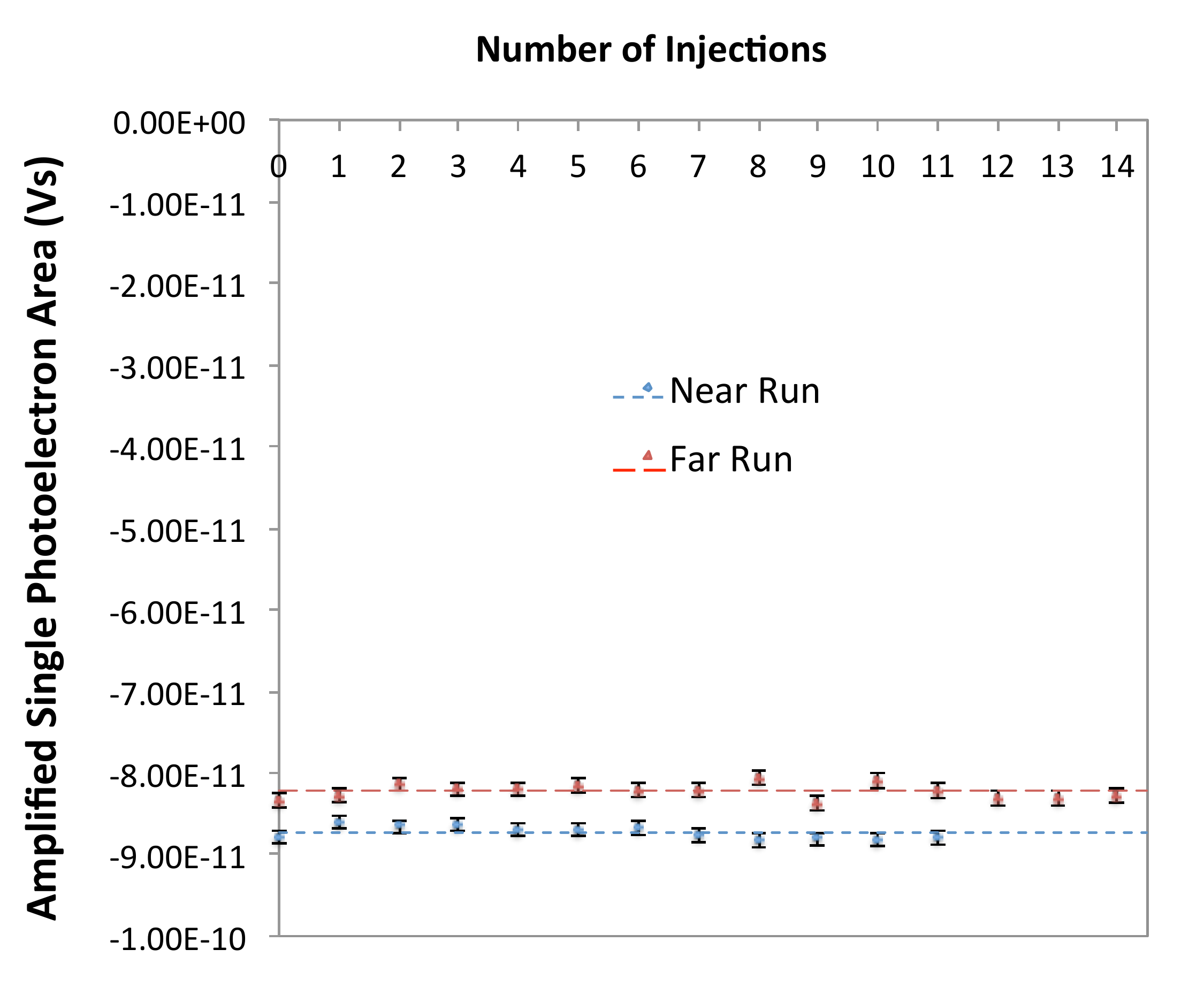}

\caption{Left: Single photoelectron sample area distribution for the pre-injection data point in the near source configuration run. Right: Single photoelectron stability for each of
the two runs reported in this paper. The dashed lines show the average
single PE size for each run, used to normalize the alpha pulses. The
error bars represent the spread of measurements about the mean, and
are folded into the nitrogen absorption measurement as a systematic error.
\label{fig:Left:-Single-photoelectron}}
\end{figure}

To measure <SPE>, A 400 nm LED is driven at a rate of 100 Hz and connected to an optical fiber which penetrates the Bo lid through a fiber-optic feedthrough.  The other end of the fiber is pointed directly at the PMT photocathode from a distance of around 1 cm.  The LED driving voltage is chosen such that a significant 
fraction of all pulses do not produce any measurable response in the PMT. 
In this way, we collect a sample of mostly single photoelectron pulses.
Producing a histogram of the
areas of PMT pulses initiated by this low intensity LED gives a
superposition of the amplified single photoelectron distribution with a sharply
falling baseline noise spectrum.  This charge distribution is shown
in Figure \ref{fig:Left:-Single-photoelectron}, left.  At much higher LED voltages,
two and three photoelectron peaks were clearly observable at multiples of
the single photoelectron size.

We measure the average single photoelectron charge by calculating
the mean pulse area between limits of -0.03 nVs (to cut off the falling
baseline noise tail) and -0.18 nVs (to cut off unavoidable two and
higher photoelectron pulses).  This data is shown in Figure \ref{fig:Left:-Single-photoelectron}, right.  Using this method we measure a gain stability of 
0.91\% for the near configuration run and 1.09\% for the far configuration run. 
Since this is comparable to the precision of the <SPE> measurement, we conclude that the PMT gain is approximately stable and that <SPE> is a constant for the duration of each run.  To account for any gain fluctuations we fold in the observed <SPE> variation throughout the run as a systematic error on all measured light yields.   The small adjustment to <SPE> caused by the contamination of rare >1PE pulses in the sample is not corrected for, since our final result will be insensitive to this shift.

For the near configuration run, as well as using an LED to measure <SPE>, we also performed measurements of the areas of single photoelectron pulses in the slow tail of the alpha scintillation light.  The benefit of this method is that a very clean sample of known single photoelectron peaks can be obtained, with negligible two and higher photoelectron contamination.  This method yielded a single photoelectron scale variation of 1.02\% between all concentration points, in good agreement with <SPE> fluctuations measured using the LED.  The absolute single photoelectron scale measured in this manner deviates from the LED <SPE> measurement by 7\%.  This deviation is much below the level where it would have an impact upon our final result, as is shown later in in section \ref{sec:FitFuncSection} and Figure \ref{fig:ShadowAndSPEStability}.

A further constraint on the gain stability is provided by repeated measurements of the light yield at the same nitrogen contamination point performed many hours apart.  These measurements will be described in section \ref{sec:Results}, and show deviations of less than 0.5\% over around eight hours.  This method constrains deviations due to both the single photoelectron scale and also the effects of any additional outgassing impurities, such as water or oxygen.

\subsection{Construction of a Fit Function Describing Signal and Background Components}
\label{sec:FitFuncSection}

\begin{figure}[tb]
\begin{centering}
\includegraphics[width=1.0\columnwidth]{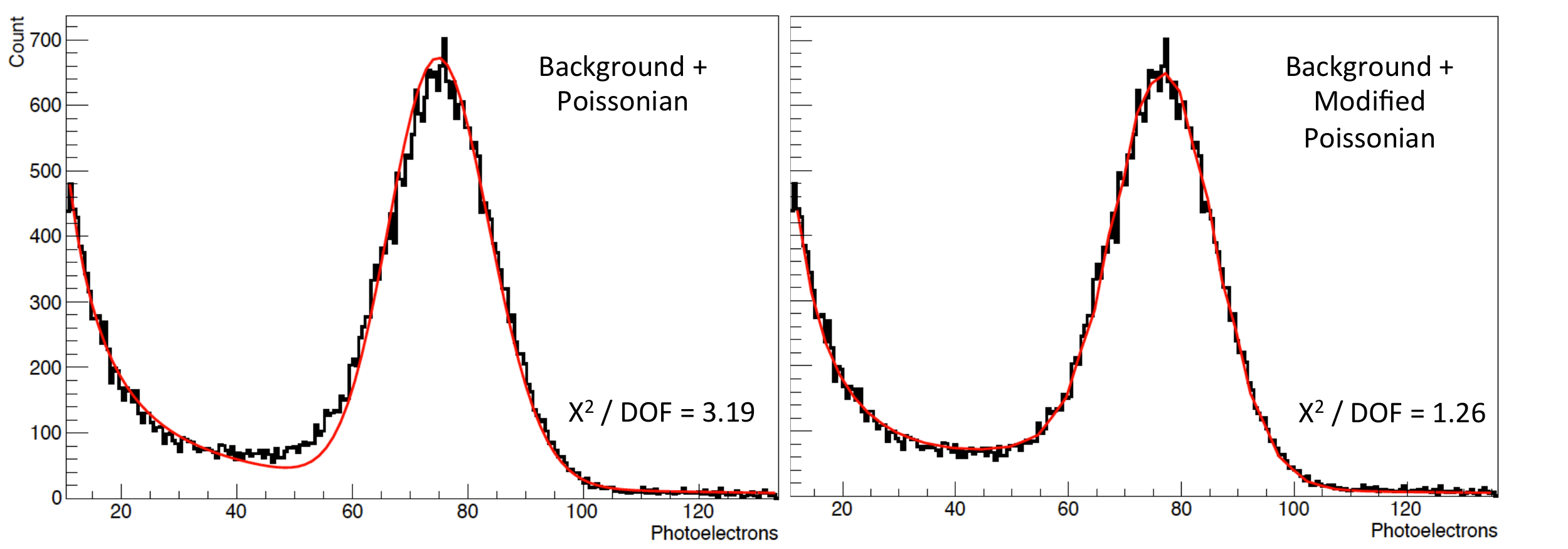}
\par\end{centering}

\caption{Left : Best fit for naive Poisson distribution plus background. Right: Best fit for modified Poisson distribution accounting for source shadowing plus background.  The shadowing function is tuned on an independent dataset, so both functions have the same number of free parameters in this fit.
\label{fig:FitFunctionBadGood}}
\end{figure}

The distributions of prompt pulse areas shown in Figure \ref{fig:SeveralDistributions} have two components: a cosmic-ray-induced falling background and an alpha-induced Poisson-like signal peak.  A simple fit to a Poisson distribution added to a power-law background model is shown in Figure \ref{fig:FitFunctionBadGood}, left, and clearly does not reproduce the measured distribution well in the region below the peak.  In this section we describe the construction of an improved function which provides a more accurate model of the detected light yield, ultimately producing the fit function shown in Figure \ref{fig:FitFunctionBadGood}, right.

There are essentially two possible explanations for the discrepancy shown in Figure  \ref{fig:FitFunctionBadGood}, left: either the cosmic-induced background is not well modeled by a power law and has an enhanced contribution in this intermediate range; or the alpha-induced pulses do not produce a simple Poisson distribution, but rather a distribution with an enhanced low tail.  Some amount of the latter effect is expected from partial occlusion of the edges of the alpha disc source by its stainless steel holder, which we refer to henceforth as "shadowing."

\begin{figure}[tb]
\begin{centering}
\includegraphics[width=1\columnwidth]{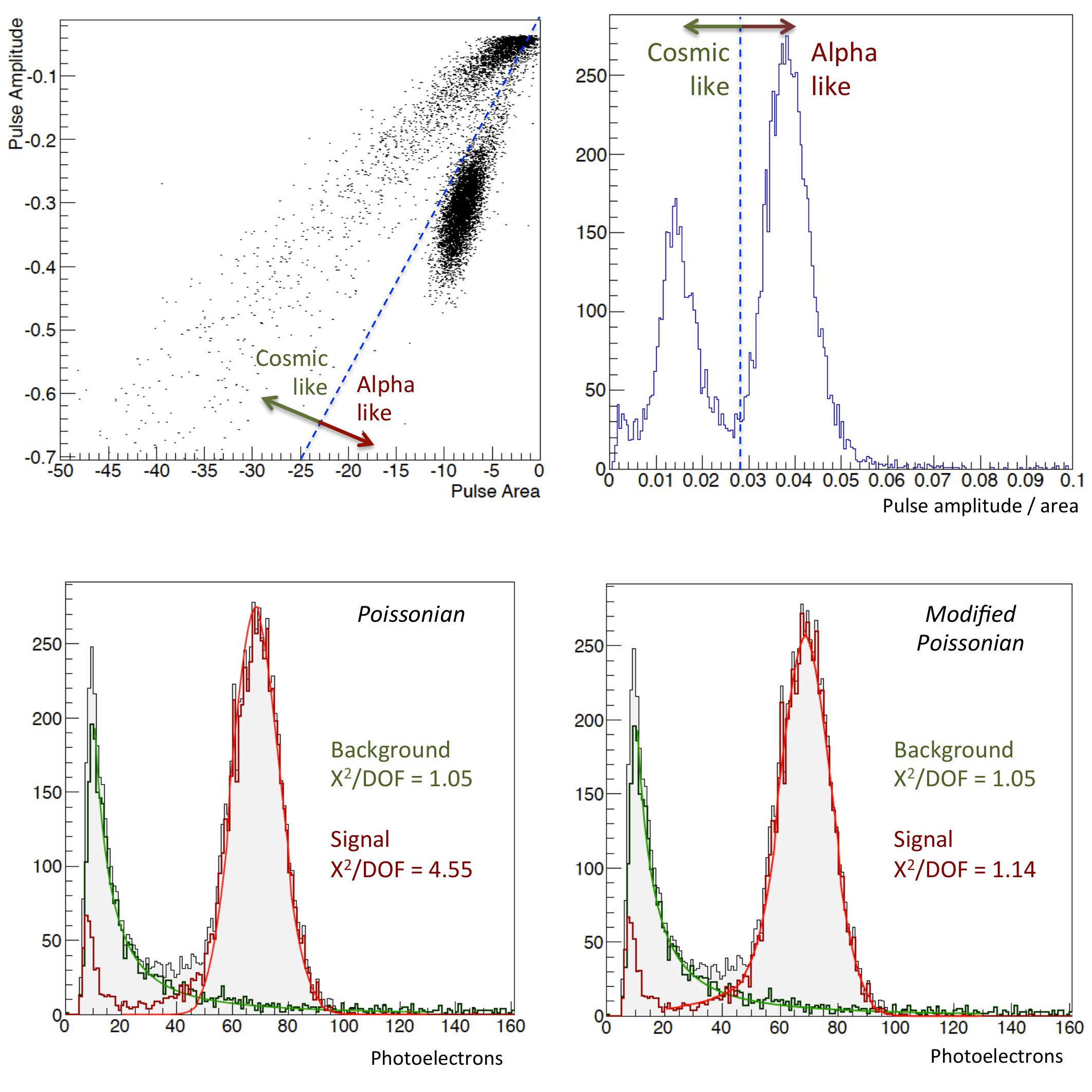}
\par\end{centering}

\caption{Top left: pulse amplitude vs area in 3.2 microseconds after the trigger for clean argon in the near configuration.  Distinct populations of alpha- and MIP-induced pulses are visible.  Top right: the ratio of total area / amplitude is used as the PSD variable.  Bottom left: fits of each component to the power-law background and Poisson distributed alpha signal model.  Bottom right: fits of each component to the power-law background and modified Poissonian model incorporating the effects of source shadowing. \label{fig:PSDPlots}}
\end{figure}

Using a sample of individual waveforms recorded in clean argon, we can separate the signal and background samples using pulse shape discrimination (PSD). Since the majority of cosmic rays are minimum ionizing particles (MIPs) whereas alpha particles are highly ionizing, the slow component of argon scintillation light is more heavily quenched for the alpha-induced subsample.  The distribution of pulse amplitudes vs total area within 3.2 microseconds of the trigger for a clean argon sample in the near configuration is shown in Figure \ref{fig:PSDPlots}, top left.  Taking the ratio of area to amplitude, two clear peaks emerge, shown in Figure \ref{fig:PSDPlots}, top right.  We separate the sample as shown in Figure \ref{fig:PSDPlots} into "cosmic-like" and "alpha-like" subsamples, with the pulse area distributions shown in the lower panels of Figure \ref{fig:PSDPlots}.  The power-law background model reproduces the cosmic-like distribution with $\chi^{2} / DOF = 1.05 $.  The low tail of the alpha-like sample is not modeled well by a Poisson distribution, suggesting some shadowing of the source by its holder is present.  Some leakage of the cosmic sample into the alpha sample is seen to occur at very low light intensity, so the region below 20 photoelectrons is not included in the alpha sample fit.  Incorporation of a shadowing function as described below improves the $\chi^{2} / DOF $ value for the alpha subsample from 4.55 to 1.14.   

We cannot rely on PSD to separate the cosmic- and alpha-induced subsamples when nitrogen contamination is present, since quenching of the argon slow component would introduce an undesirable bias into the discrimination condition.  However, since the shadowing of the source is not dependent upon the nitrogen concentration, a function describing the shadowing effect can be obtained using the clean argon sample where PSD is effective, and used to construct a modified Poisson distribution to be used for each subsequent fit.

\begin{figure}[tb]
\begin{centering}
\includegraphics[width=1\columnwidth]{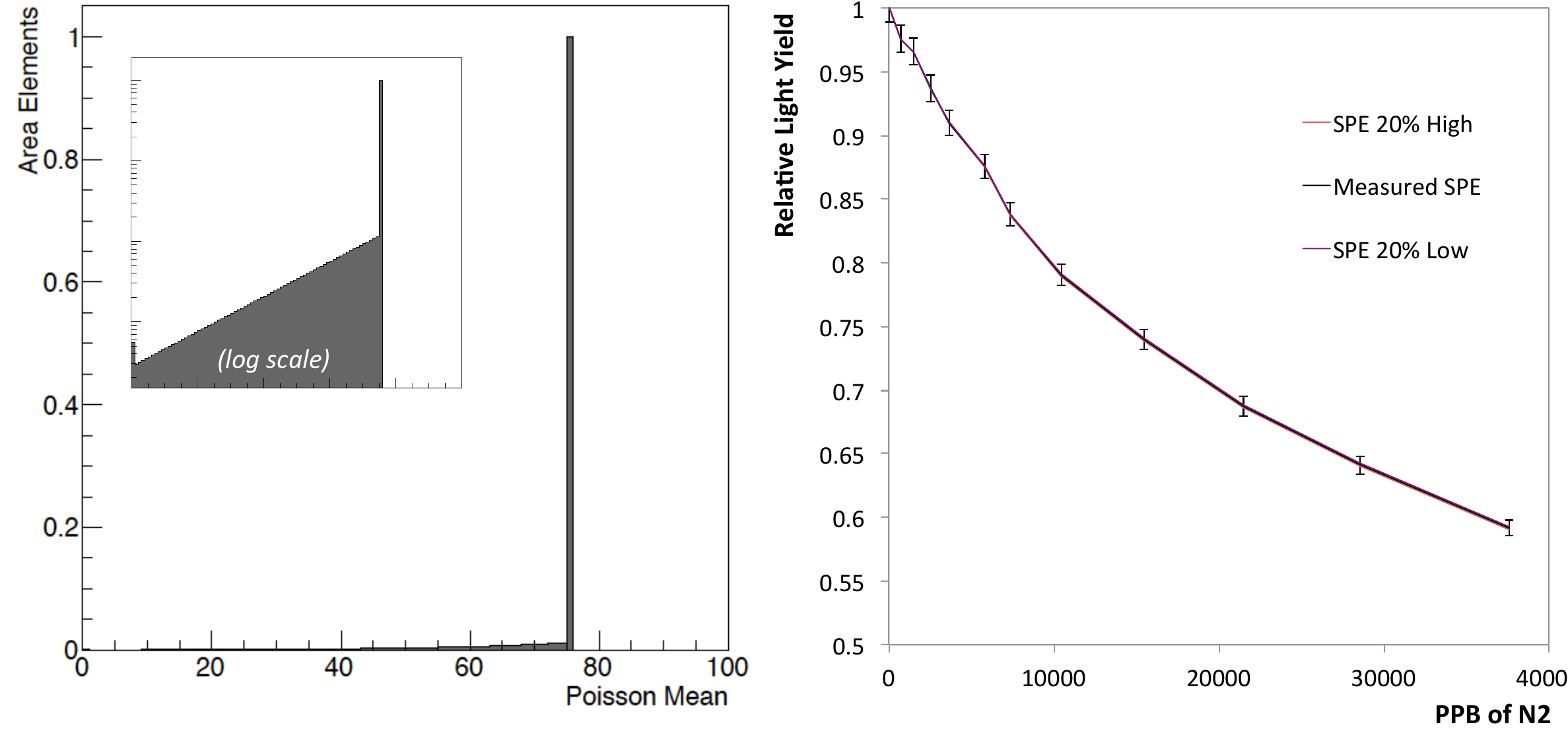}
\par\end{centering}

\caption{Left: The best-fit shadowing function for the clean argon, near-source calibration sample.  Right: the relative light yield as a function of nitrogen concentration for the near source configuration for three values of the single photoelectron scale.  \label{fig:ShadowAndSPEStability}}
\end{figure}

The form of the shadowing function is constrained by its expected scaling as the total light yield is reduced.  Due to the geometry of the source holder, most area elements on the source disk are unshadowed, all contributing with equal intensity to the main Poisson-like peak.  A small region consisting of shadowed area elements around the edge of the disk are assumed to produce Poisson-distributed light yields whose means are a fixed fraction of the Poisson mean from the unshadowed elements.  The distribution of area elements as a function of their Poisson means is referred to as the shadowing function.  The precise functional form of the shadowing function is not known a priori, but a good fit to the alpha-like distribution is given by assuming a slowly rising exponential below the unshadowed peak.  The best-fit shadowing function before convolution with Poisson distributions is shown in Figure \ref{fig:ShadowAndSPEStability}, left.  After convolution with Poisson distributions this shadowing function produces the signal distribution shown in Figure \ref{fig:PSDPlots}, bottom right. 

The final fit function used in this analysis is applied to non-PSD-separated samples and is given by a sum of two terms: a power-law-distributed background term and the modified Poisson distribution described above.  The relative light intensity is extracted from the modified Poisson distribution and normalized to the relative light intensity for clean argon, and these values are shown for the near sample in Figure \ref{fig:ShadowAndSPEStability}, right.  Since our attenuation curves are generated by taking the ratio between the contaminated-argon light yield and an initial clean-argon light yield, they should be relatively insensitive to the absolute single photoelectron scale, so long as it is stable over the course of each run.  By considering single photoelectron scales 20\% higher and 20\% lower than our measured value, we find the three measured attenuation curves are indeed indistinguishable within our systematic errors.  The three overlaid curves are all shown in Figure \ref{fig:ShadowAndSPEStability}, right. 

We note that whereas the alpha particles are nonrelativistic and do not produce Cerenkov light, the light detected from cosmic rays is expected to have a small Cerenkov component.  Using information from \cite{Antonello:2004sx} we estimate that around 1-2\% of the photons produced by cosmic muons traversing Bo originate from the Cerenkov process.  Since the absorption measurement described in this paper requires a precise measurement of the alpha light yield only, this small Cerenkov component does not affect our final result.

\section {Cross-Checks of Nitrogen Concentration in the Liquid Phase}

Even with a trace nitrogen monitor, the measurement of tiny concentrations of nitrogen in liquid 
argon has many subtleties.  To ensure that our measurements of the nitrogen concentration
in the liquid are faithful to the real concentration, we perform several cross-checks.

\begin{figure}[tb]
\begin{centering}
\includegraphics[width=1\columnwidth]{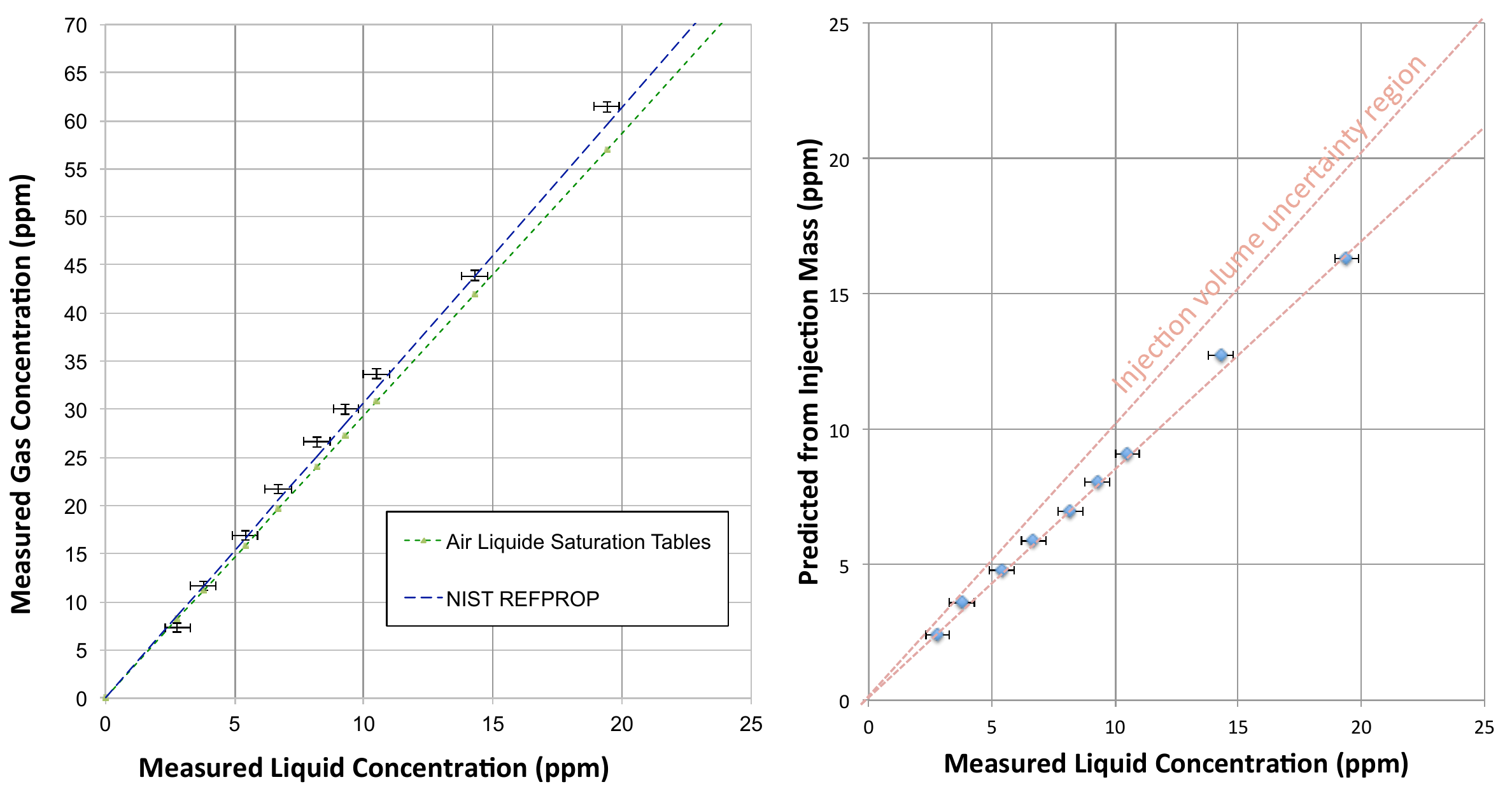}
\par\end{centering}

\caption{Cross-checks on the liquid concentration measurement. Left: measured
liquid and vapor $N_2$ concentrations compared with two calculations
of the vapor liquid equilibrium condition. Right: comparison of the
liquid $N_2$ concentration to the injected mass.  The uncertainty on the predicted injection mass comes primarily from the uncertainty of the precise injection volume.
\label{fig:Crosschecks-on-the}}
\end{figure}

After injection, nitrogen added to the cryostat becomes distributed between 
the liquid and gas phases in a vapor-liquid equilibrium, with an equilibration time of around 30 minutes.
We calculate the relationship between the equilibrium concentrations in liquid and vapor phases in two ways:  1) by referencing saturation pressure tables of argon
and nitrogen \cite{AirLiquideTables}, and 2) using the commercially available REFPROP
software package  \cite{Refprop}. By sampling from both the liquid and gas phase capillaries, 
we can determine whether the relative measured concentrations are in agreement with these predictions.
These measurements and both sets of predictions are shown in Figure \ref{fig:Crosschecks-on-the}, left.  
The agreement between the measured data points and both models indicates that 
vapor-liquid equilibrium has been reached, and that the measurement of the 
nitrogen concentration from the liquid phase sample line is indeed a faithful 
measurement of the nitrogen concentration in the liquid.

We can also predict
the expected nitrogen concentration in the liquid after each set of injections from
our knowledge of the injection volume and pressure. Due to the relevant density 
difference, the total mass of argon in the
vapor phase is much smaller than the total mass of argon in the liquid phase.  
Since the nitrogen concentration in both phases is in the few ppm range, 
the total fraction of injected nitrogen which resides in argon vapor volume is negligible
after equilibrition. Therefore, approximately all injected
nitrogen gas dissolves into the liquid. The prediction of the injected
mass of nitrogen has a large systematic error due to our uncertainty of the
exact injection volume including space in valves, capillary piping,
etc.; the limited precision of the injection pressure gauge; and the
imprecisely known temperature of the injected gas. However, the measured nitrogen
concentration in the liquid agrees with our prediction to within its
systematic error bar of 15\%, providing further support that the trace
nitrogen concentration measured from the liquid sampling capillary
is faithful to the actual nitrogen concentration in the argon volume.
A comparison between the predicted and measured concentrations is shown
in Figure \ref{fig:Crosschecks-on-the}, right. 

\section{Results}
\label{sec:Results}

\begin{figure}[tb]
\begin{centering}
\includegraphics[height=7cm]{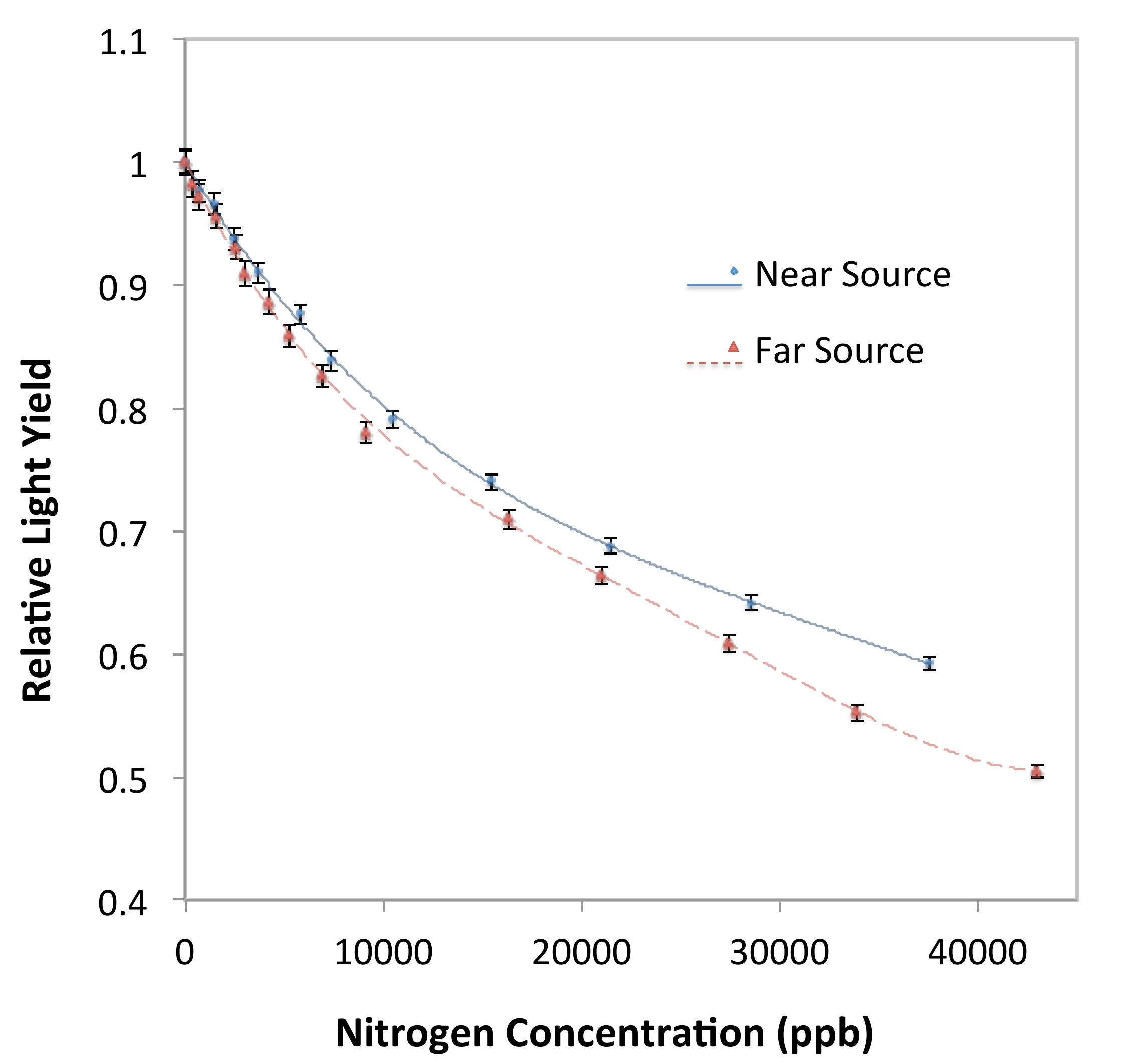}\hspace{0.2cm}\includegraphics[height=7cm]{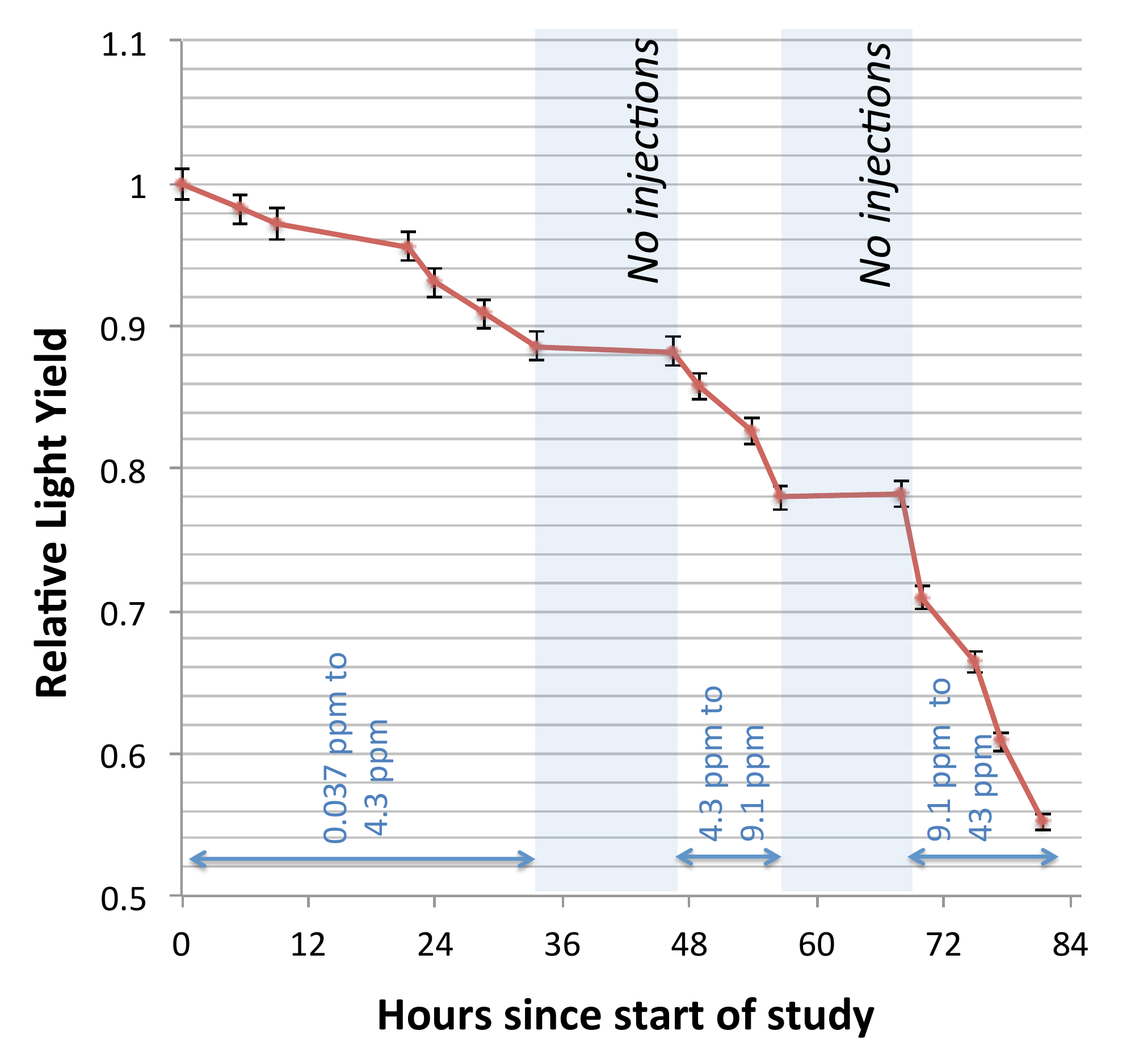}
\par\end{centering}

\caption{Left: Near and far configuration datasets, along with quartic interpolated
curves. The separation of the two curves is evidence of nitrogen
absorption. Right: The relative light yield from the far run as a function of time. During the
periods when no nitrogen was injected, the light yield remained stable to within 0.5\%. \label{fig:Left:-Near-and}}

\end{figure}

The relative light yields for each source configuration are shown
in Figure \ref{fig:Left:-Near-and}, left. A divergence of the
two datasets is visible, showing clearly the effects of nitrogen absorption.

For each concentration point we make one or more nitrogen gas injections, allow time for equilibration, record a sample of single photoelectron pulses, and then measure the alpha light yield distribution.  This process takes two to three hours, and one run of measurements takes around one week (not including time taken to purge, fill and configure the system).  To test that the observed reduction in light yield is indeed a consequence of the introduced nitrogen contamination and not the outgassing of water or some other system instability, we twice repeated measurements at the same concentration point at the end of one day and beginning of the following day, mid-run.  We found that the light yield was stable to within 0.5\% over these periods,  in contrast to its reduction during periods of nitrogen injection. This data is shown in Figure \ref{fig:Left:-Near-and}, right. 

\begin{figure}[tb]
\begin{centering}
\includegraphics[width=0.6\columnwidth]{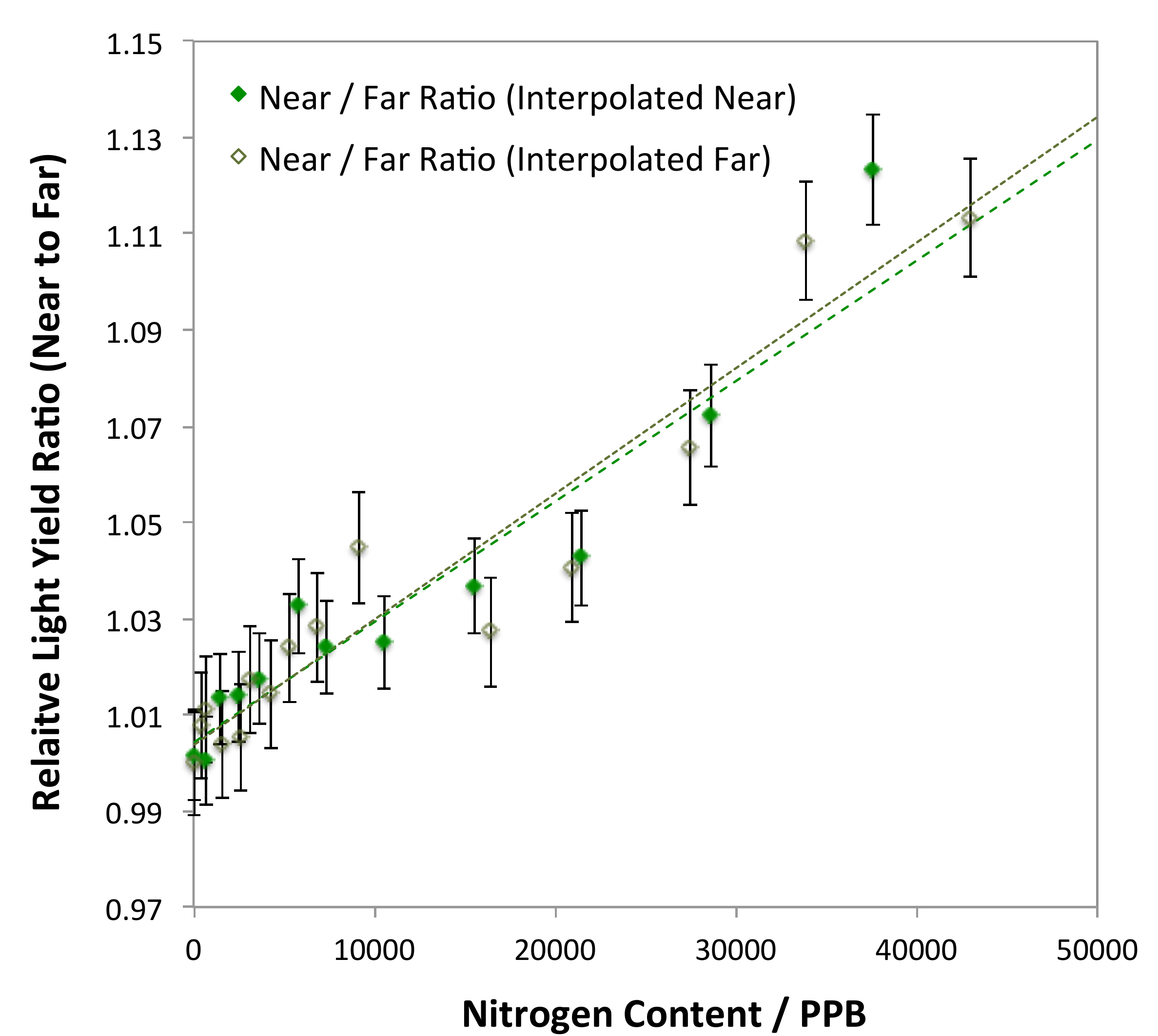}
\par\end{centering}

\caption{The fractional light yield ratio between the near and far configurations for both interpolation schemes. The dashed lines show the best linear fit for each scheme. \label{fig:RatioNearFar}}
\end{figure}

Since the precise concentration points probed in the two datasets
are different, we cannot take the ratio of light yields directly. To
find the fractional light yield ratio, we interpolate one dataset with a
polynomial and take the ratio to the measured data points of the other. 
A quartic polynomial provides a good fit to both datasets and is used
in this analysis.
We are free to choose to interpolate either the near or far dataset.
We use both schemes and then take the average of the two results
as our final value for the absorption strength.
The near-to-far ratios extracted by both interpolation schemes are
shown in Figure \ref{fig:RatioNearFar}, along with best-fit
lines for both schemes. The error bars in both cases have 
contributions from the fit error on the alpha light yield parameter
and the single photoelectron scale added in quadrature.

The largest fractional light yield ratio observed is 1.11 at 43 ppm, which, according to
our ray tracing model, corresponds to less than 1\% 
absorption per
centimeter. In this range, the relationship between the light yield
ratio and the percentage absorption per centimeter is well approximated
by a straight line with a gradient of 0.170 in units of ratio change
per (\% / cm). We can thus extract the correspondence between concentration
of nitrogen and percentage absorption per centimeter by comparing
this gradient to that of the best-fit lines of Figure \ref{fig:RatioNearFar}.

Having extracted the absorption strength in units of percentage loss
per centimeter per  ppm of nitrogen, we can further use the known density
of liquid argon to calculate the number density of nitrogen molecules per ppm 
and so extract the molecular
absorption cross section in units of $\mathrm{cm^{2}~molecule^{-1}}$. These
quantities represent the central result of this work; they are presented
in Table \ref{tab:CrossSectTable}. We take the average of the numbers from the two
interpolation schemes as our final result. The spread of these two values from their mean 
gives a measure of the systematic error introduced by using this interpolation method, 
and this is added in quadrature with the characteristic uncertainty of each measurement to
give our final uncertainty estimate.

\begin{table}[h]
\begin{centering}
\begin{tabular}{|c|c|c|}
\hline 
\textbf{Interpolation Scheme} & \textbf{Measured Absorption Strength } & \textbf{Measured Cross Section }\\& $\mathrm{cm^{-1} ppm^{-1}}$ & $\mathrm{cm^{2}~molecule^{-1}}$\tabularnewline
\hline 
\hline 
Interpolated Near & $\left(1.47\pm0.15\right)\times10^{-4}$ & $\left(6.98\pm0.71\right)\times10^{-21}$\tabularnewline
\hline 
Interpolated Far & $\left(1.54\pm0.15\right)\times10^{-4}$ & $\left(7.30\pm0.73\right)\times10^{-21}$\tabularnewline
\hline 
\textbf{\emph{Average Value}} & $\mathbf{\left(1.51\pm0.15\right)\times10^{-4}}$ & $\mathbf{\left(7.14\pm0.74\right)\times10^{-21}}$\tabularnewline
\hline 
\end{tabular}
\par\end{centering}
\caption{The absorption strength and molecular absorption cross sections of nitrogen
to argon scintillation light, as measured in this study.  Note that the near interpolation and far interpolation schemes do not give independent results, and the quoted errors are 100\% correlated. \label{tab:CrossSectTable}}
\end{table}

\section{Comparison to World Data and Implications for LArTPCs }

We have measured the absorption cross section at the wavelength of liquid argon scintillation light
of small quantities of nitrogen dissolved in liquid argon. 
This is an important parameter for the design and operation
of large LArTPCs, and in the following section we briefly
compare our result to world data on nitrogen absorption and discuss
the consequences of our measurement for these experiments.

\begin{figure}[tb]
\begin{centering}
\includegraphics[width=1\columnwidth]{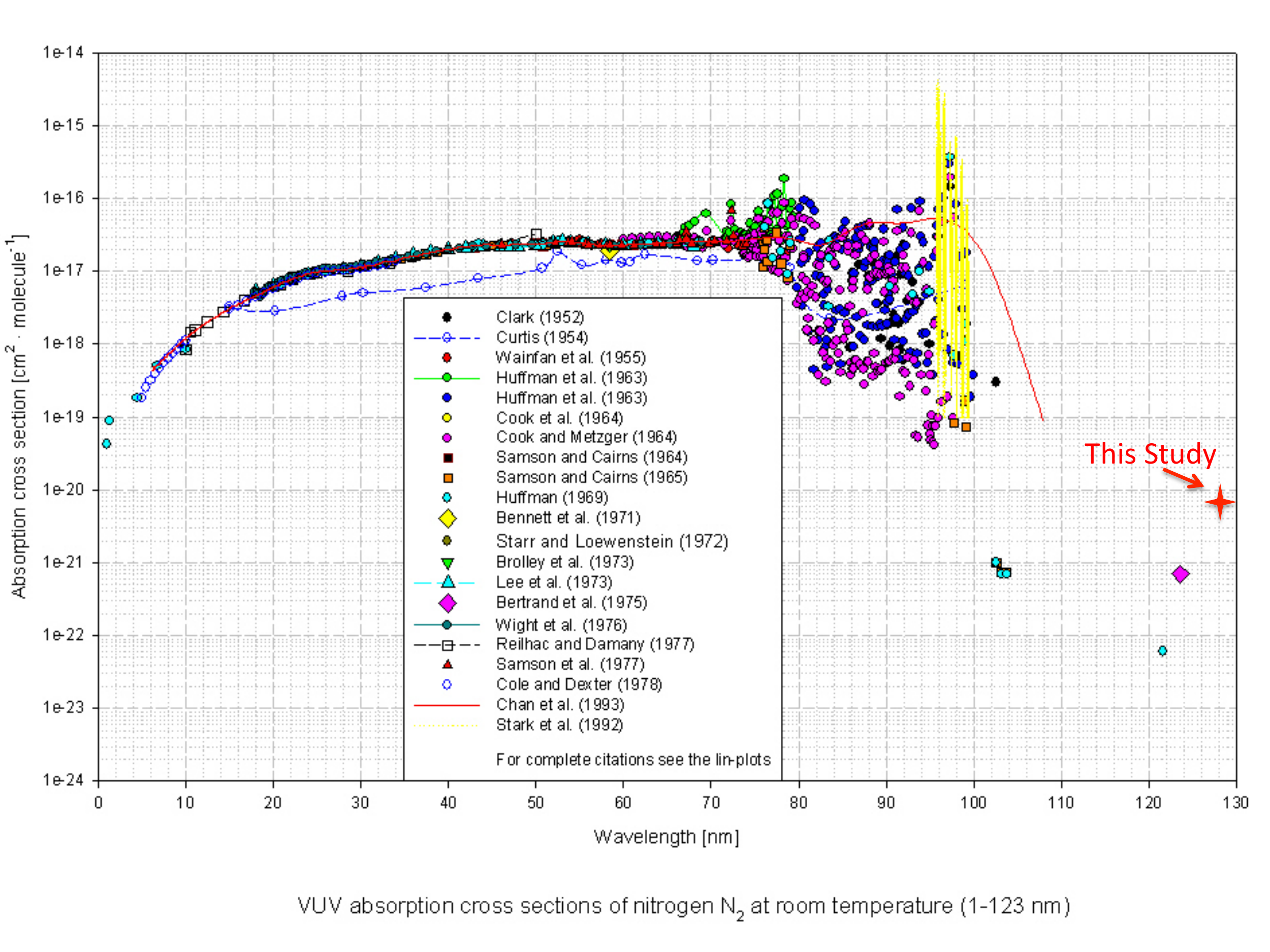}
\par\end{centering}

\caption{World data on VUV nitrogen absorption reproduced from \protect{\cite{WorldUVData}}, with
this measurement overlaid. \label{fig:World-data-on}}

\end{figure}

Figure \ref{fig:World-data-on}, reproduced from\cite{WorldUVData}, shows the world data on nitrogen absorption cross sections
in the vacuum ultraviolet (VUV) with our measurement overlaid. It is immediately clear
that measurements in the relevant wavelength range are sparse. Further,
most existing measurements consider warm nitrogen in the gas phase.
This is a very different regime from our study, which considers ppm concentrations of
 nitrogen in solution at liquid argon temperature and 10 psi above 
atmopsheric pressure.  However, despite 
these differences, our data are compatible with existing measurements and 
suggest that the nitrogen absorption cross section rises sharply 
over the range 120-130 nm.

\begin{figure}[tb]
\begin{centering}
\includegraphics[width=0.75\columnwidth]{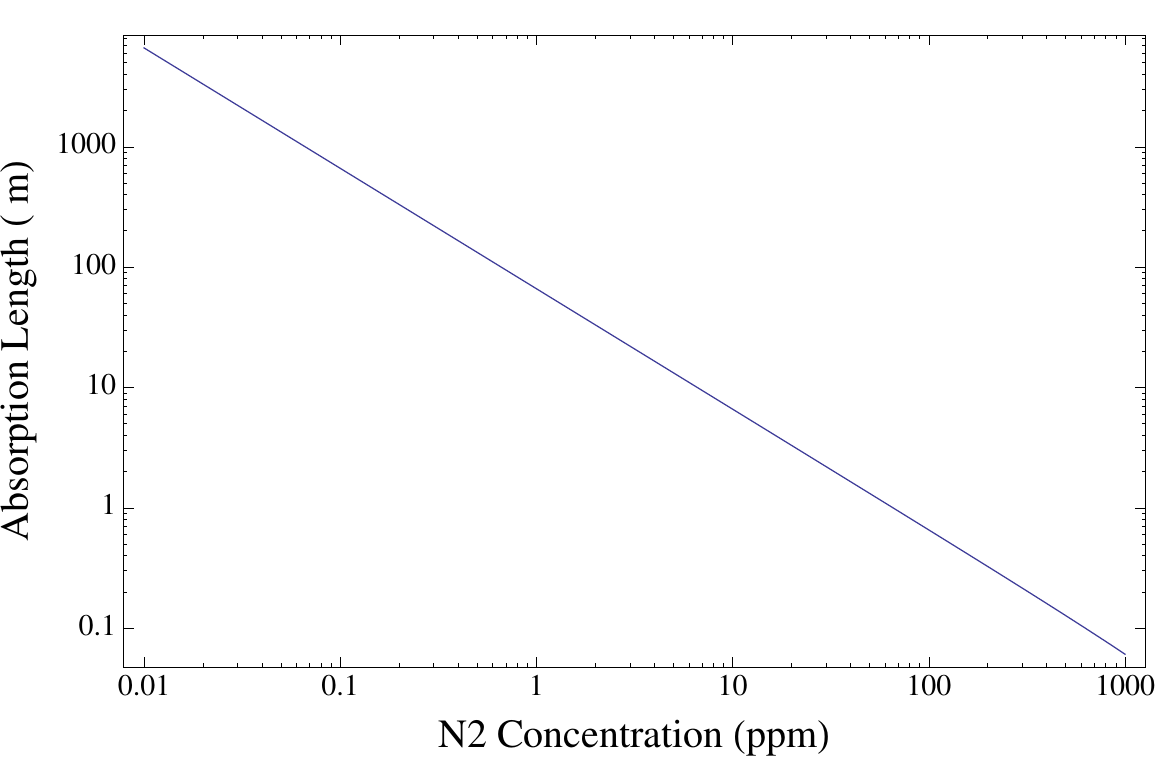}\\ \includegraphics[width=0.75\columnwidth]{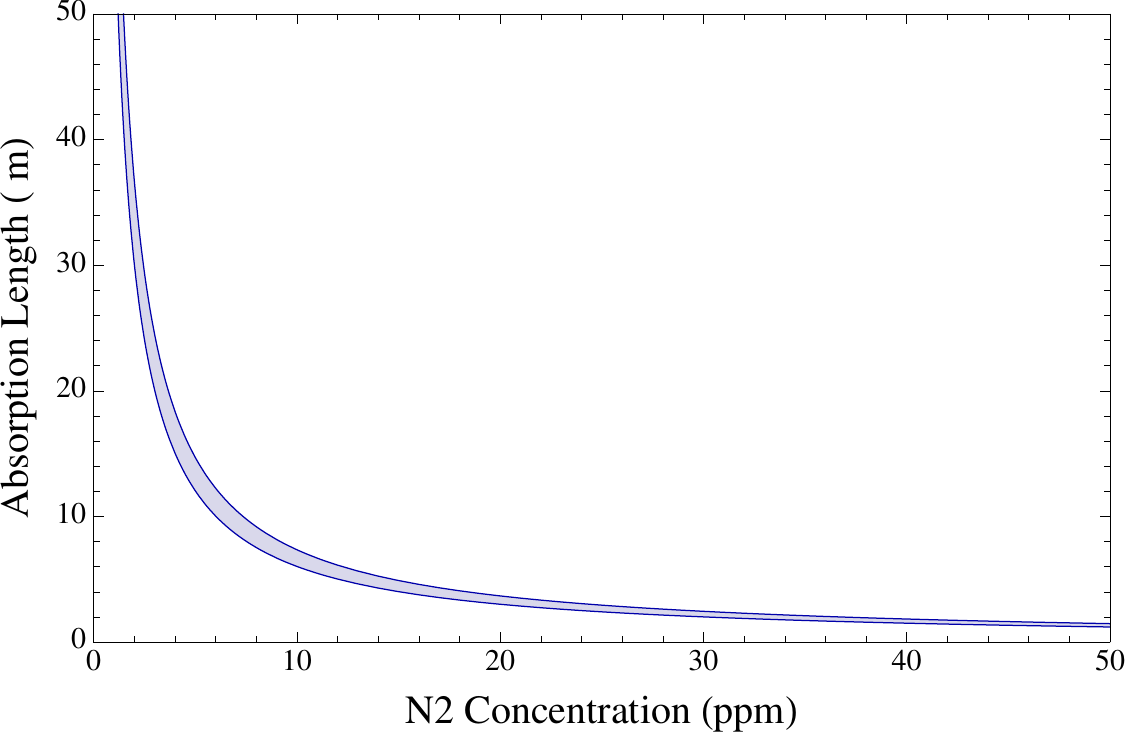}
\par\end{centering}

\caption{Absorption length of nitrogen impurities as a function of nitrogen
concentration. Top: on a log-log scale over many orders of magnitude.
Bottom: on a linear scale over the 0-50 ppm range relevant for current
generation LArTPC detectors, with the one sigma region measured in
this paper shaded. \label{fig:Absorption-length-of}}

\end{figure}

Typically the effects of impurity absorption in LArTPCs
are described in terms of an absorption length. To convert from percentage
absorption per  ppm centimeter ($p$) to an absorption length ($A$) in meters,
we use the formula

\[
A=-\frac{1m}{100\log\left(1-p\chi*1cm\right)}
\]

where $\chi$ is the concentration of nitrogen in ppm. Using the measurement
of $p$ obtained from this study, we can make
a plot of $A$ as a function of the nitrogen concentration $\chi$. We
show this relationship on a log-log plot in Figure \ref{fig:Absorption-length-of}
top, and in the region of interest from 0 to 50 ppm on a linear plot
in Figure \ref{fig:Absorption-length-of}, bottom. Table \ref{tab:DifferentArgonsTable} lists a few useful reference points
of nitrogen concentration in different liquid argon samples.

\begin{table}[tb]
\begin{centering}
\begin{tabular}{|c|c|c|}
\hline 
\textbf{Argon Specification} & \textbf{Concentration of $\mathrm{N_2}$}  & \textbf{ Absorption Length } \tabularnewline
\hline 
\hline 
Measured $N_2$ concentration of  &  37 ppb  &  1790 $\pm$ 160  m  \\ clean argon for this study & & \tabularnewline
\hline 
AirGas research (grade 6) argon \cite{AirGasSpec} &  1 ppm  &  66 $\pm$ 6 m      \tabularnewline
\hline 
MicroBooNE cryogenic specification &  2 ppm  &  30 $\pm$ 3 m      \tabularnewline
\hline 
Start of liquid recirculation phase of    &  8 ppm  &  8.2 $\pm$ 0.7 m   \\   Liquid Argon Purity Demonstrator, Run 2 \cite{Rebel:2011zzb, LAPDTalk} & & \tabularnewline
\hline 
AirGas industrial (grade 4) argon \cite{AirGasSpec} &  100 ppm  &  0.65 $\pm$ 0.06 m      \tabularnewline
\hline 
\end{tabular}
\par\end{centering}
\caption{The absorption length expected in argon samples of different nitrogen contamination levels.  These values assume that absorption due to other impurities such as water and oxygen, and the self absorption of liquid argon, can be neglected in all cases.  This is likely to be a good assumption for all but the first entry. \label{tab:DifferentArgonsTable}}
\end{table}

As well as being a vital parameter for the design of cryogenic systems for large LArTPCs and possible future dark matter experiments, the absorbing effects
of nitrogen must also be taken into account for simulation and data analysis of the current generation of experiments. As a representative example, in the MicroBooNE experiment the concentration of nitrogen in the liquid will be monitored using a trace nitrogen analyzer, similar to that used in this study, at regular intervals.  Using the attenuation length vs. concentration relationship measured here, the effects of attenuation due to nitrogen can be incorporated into optical simulations, and corrections may be applied to optical data either before or after event reconstruction.  This is likely to be particularly important for the optical detection of low-energy processes such as supernova neutrinos.

However, the main conclusion which can be derived from this study is that the absorbing effect of nitrogen to argon scintillation light is relatively weak.  Even with a loose purity specification of 10 ppm of nitrogen, the absorption length of argon with dissolved nitrogen is several meters.  Hence the measurement presented in this paper suggests that nitrogen absorption is unlikely to be a significant problem for large LArTPCs, assuming certain relatively weak purity constraints are satisfied.  This information may help to reduce the costs involved in designing and filling such future detectors.

\newpage
\acknowledgments

We thank Stephen Pordes for giving us access to the Bo cryostat and its associated cryogenic system, and for a careful reading of this manuscript.  We also thank Jong Hee Yoo and Adam Para for kindly
loaning us pieces of equipment used in this study, Clementine Jones for proofreading 
this paper and Roberto Acciarri and Flavio Cavanna for offering helpful guidance and insightful comments. 
We are very grateful
to Bill Miner, Ron Davis and the other technicians who have assisted us 
at the Proton Assembly Building,
Fermilab for their tireless hard work to provide us with cryogenic
facilities of the very highest standard. The authors thank the National
Science Foundation (NSF-PHY-1205175) and Department Of Energy (DE-FG02-91ER40661).
This work was supported by the Fermi National Accelerator Laboratory,
which is operated by the Fermi Research Alliance, LLC under Contract
No. De-AC02- 07CH11359 with the United States Department of Energy.

\bibliographystyle{JHEP}
\bibliography{N2AbsorptionPaper.bib}{}

\end{document}